\documentclass{aa}  
\usepackage{graphicx}
\usepackage{txfonts}
\usepackage{color}
 \usepackage[normalem]{ulem} 

\usepackage{natbib,twoopt}
\usepackage[breaklinks=true]{hyperref} 
\bibpunct{(}{)}{;}{a}{}{,}             
  \newcommandtwoopt{\citeads}[3][][]{\href{http://adsabs.harvard.edu/abs/#3}%
    {\def\hyper@linkstart##1##2{}%
     \let\hyper@linkend\@empty\citealp[#1][#2]{#3}}}
  \newcommandtwoopt{\citepads}[3][][]{\href{http://adsabs.harvard.edu/abs/#3}%
    {\def\hyper@linkstart##1##2{}%
     \let\hyper@linkend\@empty\citep[#1][#2]{#3}}}
  \newcommandtwoopt{\citetads}[3][][]{\href{http://adsabs.harvard.edu/abs/#3}%
    {\def\hyper@linkstart##1##2{}%
     \let\hyper@linkend\@empty\citet[#1][#2]{#3}}}
  \newcommandtwoopt{\citeyearads}[3][][]%
    {\href{http://adsabs.harvard.edu/abs/#3}
    {\def\hyper@linkstart##1##2{}%
     \let\hyper@linkend\@empty\citeyear[#1][#2]{#3}}}
\makeatother
\usepackage{todonotes}

\def\la{\mathrel{\hbox{\rlap{\hbox{\lower4pt\hbox{$\sim$}}}\hbox{$<$}}}}
\def\ga{\mathrel{\hbox{\rlap{\hbox{\lower4pt\hbox{$\sim$}}}\hbox{$>$}}}}

\def\deg      {{\ifmmode^\circ\else$^\circ$\fi} } 
\def\arcmin   {{\ifmmode {'}\else$'$\fi}} 
\def\arcsec   {{\ifmmode{''}\else$''$\fi}} 

\begin{document} 
   \title{A disturbance in the force}
    \subtitle{How force fluctuations hinder dynamical friction and induce core stalling}
   \author{Pierfrancesco Di Cintio
          \inst{1,2,3}
          \and
          Bruno Marcos
          \inst{4,5}
          }
   \institute{Consiglio Nazionale delle Ricerche, Istituto dei Sistemi Complessi, via Madonna del Piano 17, 50022 Sesto Fiorentino (FI), Italy\\
              \email{pierfrancesco.dicintio@cnr.it}
         \and
             INFN -- Sezione di Firenze, via G. Sansone 1, 50022 Sesto Fiorentino (FI), Italy
         \and
             INAF -- Osservatorio Astrofisico di Arcetri, Largo E. Fermi 5, 50125 Firenze, Italy   
         \and
             Laboratoire J.-A. Dieudonn\'e, UMR 6621, Universit\'e C\^ote d'Azur, Parc Valrose 06108 Nice Cedex 02, France
         \and
             Departamento de Estructura de la Materia, F\'{i}sica T\'{e}rmica y Electr\'{o}nica, Universidad Complutense Madrid, 28040 Madrid, Spain
            \email{bruno.marcos@univ-cotedazur.fr}
             }

   \date{Received May 7, 2025; accepted xxxx xx, xxxx}
  \abstract
   {Dynamical friction is an important phenomenon in stellar dynamics (and plasma physics) resulting in the slowing down of a massive test particle upon many two-body scatters with background particles. Chandrasekhar's original formulation, developed for idealized infinite and homogeneous systems, has been found to be sufficiently accurate even in models of finite extent and radially dependent density profiles. However, in some cases $N-$body simulations evidenced a breakdown of Chandrasekhar's formalism. In particular, in the case of cored stellar systems, the analytical predictions underestimate the rate of in-fall of the test particle. Moreover, the orbital decay appears to stop at a critical radius where in principle the effect of dynamical friction should still be non-negligible.}
   {Several explanations for such discrepancy have been proposed so far, in spite of this it remains unclear whether the origin is a finite N effect or an effect arising from the resonance (or near-resonance) of the orbits of the test and field particles, which is independent on $N$, such as dynamical buoyancy. Here we aim at shedding some light on this issue with tailored numerical experiments.}
   {We perform ad hoc simulations of a massive tracer initially placed on a low eccentricity orbit in spherical equilibrium models with increasing resolution. We use an $N-$body code where the self-consistent interaction among the background particles can be substituted with the effect of the static smooth potential of the system's continuum limit, so that the higher order contributions to the dynamical friction arising from the formation of a wake can be neglected if needed.}
   {We find that, contrary to what reported in the previous literature, a suppression of dynamical friction happens in both cuspy and cored models. When neglecting the interaction among field particles we observe in both cases a clear $N^{-1/2}$ scaling of the radius at which dynamical friction ceases to be effective. This hints towards a granularity-induced origin of the so-called core-stalling of the massive tracer in cored models.}
   {}

   \maketitle
\section{Introduction}
Dynamical friction (hereafter DF, \citealt{1943ApJ....97..255C,1943ApJ....97..263C,1943ApJ....98...54C}) is the process that drives a massive particle  of mass $m_t$ (e.g. a compact star/black hole or a satellite dwarf galaxy) to sink into the host stellar system via multiple dynamical collisions with its constituting field particles of (mean) mass $m$. In its original formulation, evaluated for an infinitely extended system of constant number and mass density $n$ and $\rho=mn$ with isotropic velocity distribution $f(v)$, DF amounts to a velocity dependent drag term of the form
\begin{equation}\label{simpleDF}
\frac{{\rm d}\mathbf{v}}{{\rm d}t}=-16\pi^2G^2\rho(m_t+m)\ln\Lambda\frac{\Xi(v)}{v^3}\mathbf{v}.
\end{equation}
In the equation above, $\ln\Lambda$ is the (velocity averaged) Coulomb logarithm of the maximum to minimum impact parameters $\Lambda=b_{\rm max}/b_{\rm min}$ in the hyperbolic scattering with $1/r^2$ forces, $v=||\mathbf{v}||$, and
\begin{equation}\label{velfunct}
\Xi(v)=\int_0^{v}f(v^\prime)v^{\prime 2}{\rm d}v^\prime,
\end{equation}
is the fractional velocity volume function (e.g. see \citealt{2021isd..book.....C}) that counts the portion of field particles moving slower than $m_t$ and contributing to DF.\\
\indent Since its original inception, Equation (\ref{simpleDF}) has been applied with success to a wide plethora of astrophysical system ranging from massive stars in globular clusters (GCs) to satellite galaxies in dark matter (DM) halos. However, numerical experiments have so far revealed that a naive correction to the Chandrasekhar expression, where essentially one substitutes the local values of $\rho(r)$ and $f(v)$ in Eq. (\ref{simpleDF}), might lead to either an overestimation or underestimation of DF. For example, in a series of papers also accounting for the feedback of $m_t$ on the host system by \cite{2003A&A...405...73B,2006A&A...453....9A,2007A&A...463..921A}, it has been shown that the classical idealized DF formulation works rather well in models with broad density cores, while it fails (i.e. predicts shorter sinking time-scales) when the density profile is highly concentrated at small radii. In addition, in the latter case the feedback on the density distribution is less and less important the higher is its initial concentration. Conversely, \cite{2006MNRAS.373.1451R} and \cite{2006MNRAS.368.1073G,2010ApJ...725.1707G} observed that massive objects moving on initially circular orbits in cored models first suffer a stronger DF, with respect to the Chandrasekhar estimate -often referred to as super-Chandrasekhar- to then ''stall" at a radius $r_s$ that appears to depend on specific details of the galaxy model and on $m_t$. This apparent suppression of DF\footnote{Models with the same density profile $\rho$ supported by different velocity distribution distributions can exert considerably different friction on the same mass with analogous orbital parameters. In particular, DF in radially anisotropic Osipkov-Merritt models has been found to be less effective than in the associated isotropic system (see Appendix B in \citealt{2025PhRvD.111f3071K}).} in stellar (or DM) cores, frequently dubbed core-stalling, has been also reported among the others by \cite{2016MNRAS.455.3597Z} for satellites sinking in flat galactic cores, by \cite{2018ApJ...868..134K}, \cite{2019ApJ...877..133D} and \cite{2012MNRAS.426..601C,2024A&A...690A.119B} in simulations of GCs orbiting dwarf galaxies, and by \cite{2024MNRAS.534..957G} in simulations of supermassive black holes (SMBHs) in nuclear star clusters.\\ 
\indent \cite{2021ApJ...912...43B,2022ApJ...926..215B} (see also \citealt{2025arXiv250523905D}) interpreted the core-stalling (at least in very weakly collisional systems) as an effect of the \cite{1972MNRAS.157....1L} torque generalized such that its original secular and adiabatic assumptions are relaxed to account for the contribution of the non-resonant orbits to the slowing down of the massive test particle. In practice once the test particle reaches the central core, the structure of the relative orbits of the near resonant field particles $m$ with respect to $m_t$ (e.g. see \citealt{2011MNRAS.416.1181I}) changes significantly. Since the phase space distribution in nearly harmonic cores is characterized by rather gentle gradients, the net torque from said orbits is statistically almost always positive, thereby countering the friction (an effect labeled as dynamical buoyancy). The core-stalling is therefore happening once the torque induced buoyancy balances DF. Remarkably, another resonance based argument (see \citealt{1984MNRAS.209..729T}) was sketched by \cite{2006MNRAS.373.1451R} but later ruled out by \cite{2015MNRAS.454.3778P,2016MNRAS.463..858P} that interpreted the so-called super-Chandrasekhar regime observed in some $N-$body simulations as consequence of using a local Maxwellian approximation to evaluate $\Xi(v)$, instead of integrating over the phase-space distribution $f$ that supports the specific density profile of the host model (see Appendix \ref{plummerfriction}, see also \citealt{2004ApJ...601...37K}).\\ 
\indent For the tightly related problem of a heavy particle (BH) initially sitting at rest at the centre of a flat-cored system then migrating outwards due to the combined action of force fluctuations and torques\footnote{We note that a substantial mitigation of the DF force on stellar bars (e.g. see \citealt{1985MNRAS.213..451W}), ascribed to the suppression of the torque acting on the bar, has been also reported recently by \citealt{2024MNRAS.528.4115C} in systems with DM halos with prograde rotation with respect to the galaxy's disk.}, $N-$body simulations (e.g. \citealt{2002PhRvL..88l1103C,2002ApJ...572..371C,2003ApJ...592...32C,2011MNRAS.418.1308B,2023A&A...673A...8D}) revealed a strong resolution dependence of the wander radius $r_{\rm wan}$ (\citealt{1976ApJ...209..214B}), loosely defined as the typical displacement of $m_t$ from the center of the potential well, in the form
\begin{equation}\label{rwan}
r_{\rm wan}\approx r_c\sqrt{\frac{m}{m_t}}=r_c\sqrt{\frac{M}{Nm_t}},
\end{equation}
where $r_c$ is the typical scale length of the model, $M$ its total mass and $N$ the number of stars/simulation particles. Another radial scale, considered in the BH dynamics that is usually compared with $r_{\rm wan}$, is the so-called influence radius (i.e. the radius below which the potential $\Phi_t=-Gm_t/r$ exerted by $m_t$ dominates over the contribution of the host stellar system, e.g. see \citealt{1972ApJ...178..371P}) defined as
\begin{equation}\label{rinflu}
    r_{\rm inf} = \frac{Gm_t}{\sigma^2}  \approx  r_c \frac{m_t}{M}=r_c\frac{m_t}{Nm},
\end{equation} 
where $\sigma$ is the velocity dispersion. This established, it remains unclear whether, at least in some regimes of $N$ and $m_t$, the core stalling is a discreteness effect linked to the rapid and incoherent force fluctuations in dense cores, or a resolution-independent process related to the instantaneous fraction of orbits with frequencies nearly resonant with the test particle $m_t$. In this work we explore this matter further by means of variable resolution $N-$body simulations and single particle propagation in smooth potentials with tunable DF and mean force fluctuations.  
\section{Setting the stage}
\subsection{Galaxy models}
In most of the simulations discussed in this paper we employ the frequently used $\gamma-$models (\citealt{1993MNRAS.265..250D,1994AJ....107..634T}) with density distribution defined by
\begin{equation}\label{dehnen}
\rho(r)=\frac{3-\gamma}{4\pi}\frac{Mr_c}{r^\gamma(r+r_c)^{4-\gamma}};\quad 0\leq\gamma<3.
\end{equation}
generating the gravitational potential
\begin{eqnarray}\label{phidehnen}
 \Phi(r)=\frac{GM}{r_c}\times
\begin{cases}
 \displaystyle\frac{1}{(2-\gamma)}\left[\left(\frac{r}{r+r_c}\right)^{2-\gamma}-1\right]\quad {\rm for}\quad\gamma\neq 2,\\
\displaystyle\ln\frac{r}{r+r_c} \quad {\rm for}\quad\gamma=2.
\end{cases} 
 \end{eqnarray}
The parameter $\gamma$ appearing in Eqs.(\ref{dehnen},\ref{phidehnen}) above controls the central logarithmic density slope, ranging from 0 (corresponding to a flat core) to 3 (corresponding to a singularity). Here we concentrate on the cored $\gamma=0$ and mildly cuspy $\gamma=1$ cases. When generating the simulations initial conditions, the radial coordinates $r_i$ of the $N$ field particles are obtained sampling $\rho$ with the usual Monte Carlo technique applied to the radial mass profile
\begin{equation}
M(r)=M\left(\frac{r}{r+r_c}\right)^{3-\gamma},
\end{equation}
 \begin{figure*}
	\centering
	\includegraphics[width=0.45\textwidth]{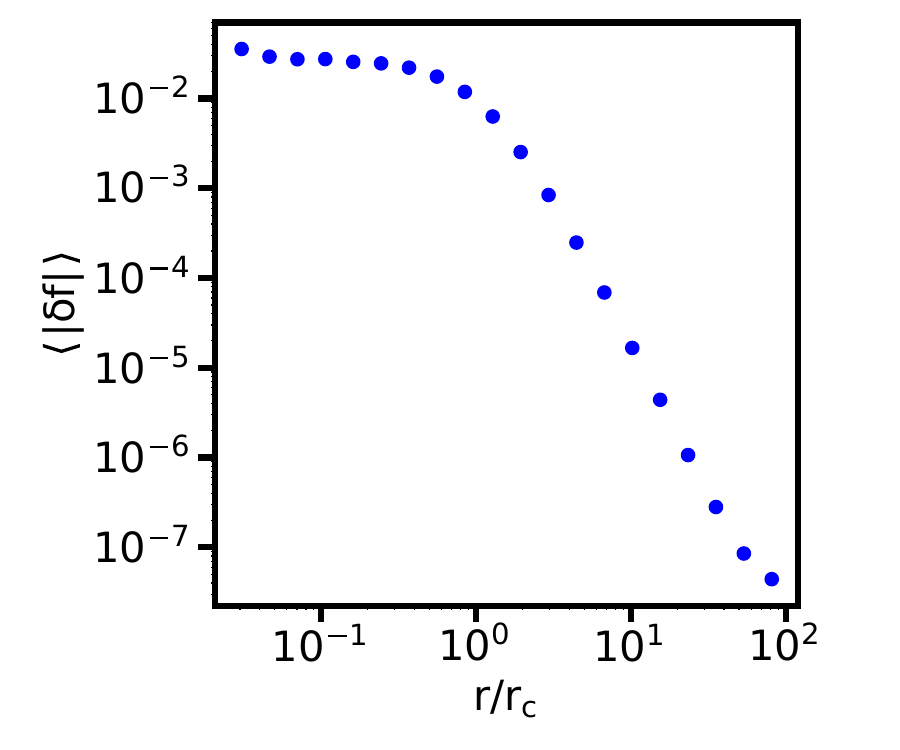}
        \includegraphics[width=0.45\textwidth]{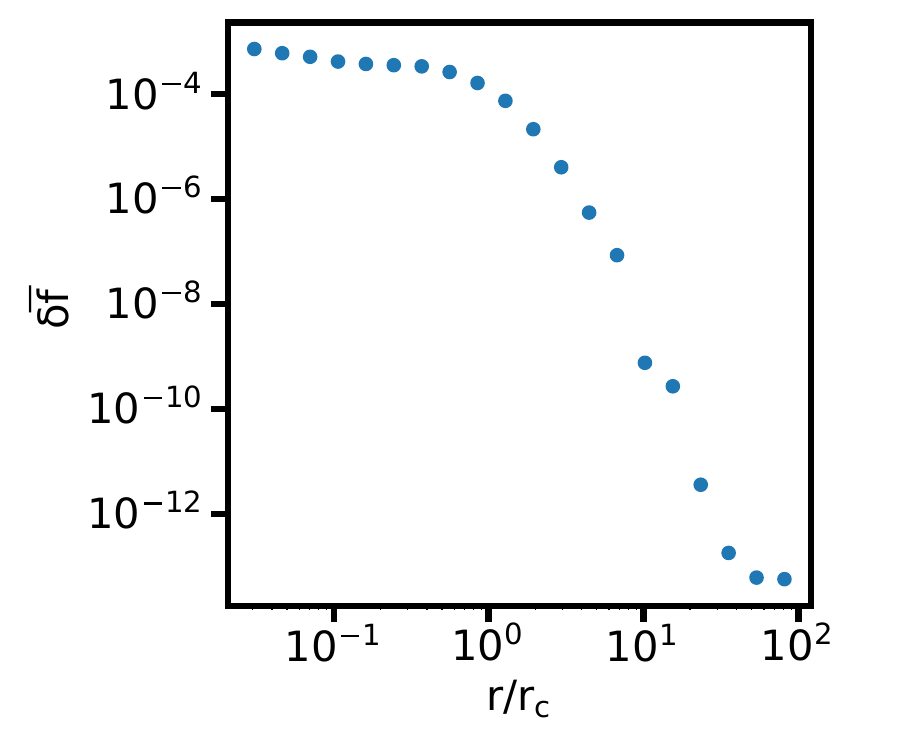}
	\caption{Radial profile of the average of the discrete force fluctuations (left panel) and its variance (right panel) evaluated for a single realization of a Plummer system with $N=10^6$.}
	\label{plumer_average_variance}
\end{figure*}
while their angular positions $\vartheta_i$ and $\varphi_i$ are assigned randomly from two independent uniform distributions. The modulus of the velocities $v_i$ are assigned using the rejection method (for example, see \citealt{2015JPlPh..81e4904D} and references therein) to sample the isotropic equilibrium phase-space distribution
\begin{equation}\label{distrfunct}
f(\mathcal{E})=\frac{(3-\gamma)M}{2(2\pi^2GMr_c)^{3/2}}\int_0^\mathcal{E}\frac{(1-y)^2}{y^{4-\gamma}\sqrt{\mathcal{E}-\Psi}}{\rm d}\Psi,
\end{equation}
where $\Psi=-\Phi$ and $\mathcal{E}=\Psi-v^2/2$ are the relative potential and energy, respectively and
\begin{equation}
\label{ypsi}
y\left(\Psi\right)=
\begin{cases}
\displaystyle \left[1-(2-\gamma)\Psi\right]^{1/(2-\gamma)},\quad \gamma\neq 2,\\
\displaystyle \exp\left(-\Psi\right),\quad \gamma=2.
\end{cases}
\end{equation}
In addition, to compare with previous numerical work, we have also considered the simpler \cite{1911MNRAS..71..460P} model defined by the density potential pair
\begin{equation}\label{plummer}
    \rho(r)=\frac{3 M}{4\pi r_c^3}\left( 1 + \frac{r^2}{r_c^2} \right)^{-5/2};\quad  \Phi(r)=-\frac{GM}{\sqrt{r^2+r_c^2}},
\end{equation}
with radial mass profile 
\begin{equation}
M(r)=M\left(\frac{r^2}{r^2+r_c^2}\right)^{3/2}.
\end{equation}
The isotropic phase-space distribution function associated to the model (\ref{plummer}) is
\begin{equation}\label{fplummer}
f(\mathcal{E})=\frac{\sqrt{2}}{378\pi^2Gr_c^2\sigma_0}\left(\frac{\mathcal{E}}{\sigma_0^2}\right)^{7/2},
\end{equation}
where $\sigma_0=\sqrt{GM/6r_c}$ is the central velocity dispersion. We note that the radial gravitational acceleration $a_r=-GM(r)/r^2$ in the Plummer model and the $\gamma=0$ model are both nearly harmonic near their centres, however for the same choice of $M$ and $r_c$, the radial span over which $a_r\propto r$ is larger for the Plummer density profile. 
\subsection{$N-$body method and simulation setup}
We have performed $N-$body simulations where a test mass $m_t$ sinks in an equilibrium system represented by $N$ equal mass $m=M/N$ field particles, with $10^4\leq N\leq 3\times 10^6$. The test particle mass values used in this work range from $2\times 10^{-4}$ to $10^{-2}$ in units of the total mass of the host system $M$, while the orbit of $m_t$ is initialized with values of the eccentricty $e$ ranging from 0 (i.e. circular orbit) to 0.9 corresponding to relatively eccentric orbit. As usual $e$ is defined as
\begin{equation}
e = \frac{r_{\text{apo}} - r_{\text{peri}}}{r_{\text{apo}} + r_{\text{peri}}},
\end{equation}
where $r_{\text{apo}}$ and $r_{\text{peri}}$ are the apocentre and pericentre of the unperturbed orbit in the smooth potential $\Phi$.\\
\indent In our numerical experiments the field particles either interact among each other, with forces computed with the fast \cite{2000ApJ...536L..39D} tree scheme implemented in the {\sc fvfps} code (see \citealt{2003MSAIS...1...18L,2006MNRAS.370..681N}), or move as independent tracers in the smooth potential given by Eqs. (\ref{phidehnen}), (\ref{plummer}) in a similar fashion to \cite{1983ApJ...274...53W} and \cite{1987MNRAS.224..349B}. In this latter case we aim at neglecting the higher order effects on DF induced by the formation of a self-gravitating wake (\citealt{1983A&A...117....9M,2022MNRAS.515..407K}) behind $m_t$, sometimes also referred to as polarization cloud (\citealt{1970ApJ...159..239G,1972ASSL...31...13K,2017MNRAS.469.4193H}). In both types of simulation, each field particle of mass $m$ interacts with the test particle $m_t$ with the gravitational force evaluated with a direct sum scheme. To prevent unwanted resolution-induced force singularities, the Newtonian $1/r^2$ interaction is substituted by a quadratic spline below the softening length $\epsilon$. In the results discussed hereafter the fixed choice was $\epsilon=2\times10^{-3}r_c$. We recall that, in direct- and tree-codes the softening act as a sort of spatial resolution. We recall also that, the introduction of said softening effectively dictates a fixed minimum impact parameter in two body-encounters (see \citealt{1993ApJ...404...73H,2019JMP....60e2901C}). All particles are propagated with a second order leapfrog scheme with fixed time step $\Delta t=t_{c}/150$, where the reference time scale is $t_c=\sqrt{r_c^3/GM}$. We adopted units such that $G=M=r_c=t_c=1$.
\subsection{Stochastic equation of motion models}
As a reference we have also performed simulations where $m_t$ is propagated in the smooth potential under the additional effect of DF and force fluctuations modeling the the ''granularity" of the system, in principle with arbitrary resolution, without resulting to integrating the field particle orbits. In practice, following the approach of \cite{2025arXiv250322479S} we assume that the dynamics of $m_t$ is statistically equivalent to that induced by the stochastic differential Langevin equation (see \citealt{1943RvMP...15....1C,1980PhR....63....1K}) 
\begin{equation}\label{langeq}
\ddot{\mathbf{r}}=-\nabla\Phi(\mathbf{r})-\eta\mathbf{v}+\delta\mathbf{f}(\mathbf{r});\quad \mathbf{v}=\dot{\mathbf{r}},
\end{equation}
where $\eta\mathbf{v}$ is essentially the DF term given in Eq. (\ref{simpleDF}) and $\delta f$ is a stochastic force (per unit mass) fluctuation. The formulation of the problem given by Eq. (\ref{langeq}) is equivalent to the evolution equation for the particle energy fluctuations $\langle{\rm d}E/{\rm d}t\rangle$ derived by \cite{1992ApJ...390...79B} and \cite{1993MNRAS.263...75M} for a homogeneous or inhomogeneous gravitational media using a fluctuation-dissipation approach.\\
\indent In the case of an infinite and homogeneous system of equal mass particles interacting via $1/r^2$ forces subjected to Poissonian density fluctuations, $\delta f$ is distributed according (see \citealt{1942ApJ....95..489C,chandra_vonnumann}) to the \cite{1919AnP...363..577H} distribution, originally evaluated for an infinite gas of charged particles. The Holtsmark distribution of the modulus of the force fluctuations $\delta f=||\delta\mathbf{f}||$ reads in integral form as function of the dimensionless auxiliary variable $s$ as
\begin{equation}\label{holtsmark}
\mathcal{F}(\delta f)=\frac{2}{\pi \delta f}\int_0^\infty\exp\left[-\alpha(s/\delta f)^{3/2}\right]s\sin(s){\rm d}s,
\end{equation}
while its density dependent normalization factor $\alpha$ is defined as
\begin{equation}
\alpha=\frac{4}{15}(2\pi G)^{3/2}m^{1/2}\rho.
\end{equation}
\begin{figure*}
	  \centering
\includegraphics[width=0.33\textwidth]{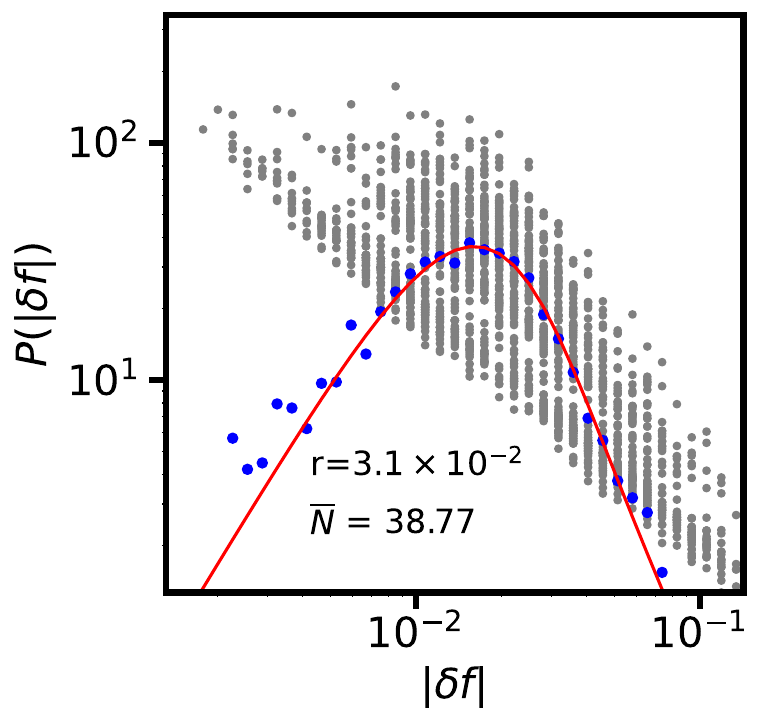}
\includegraphics[width=0.33\textwidth]{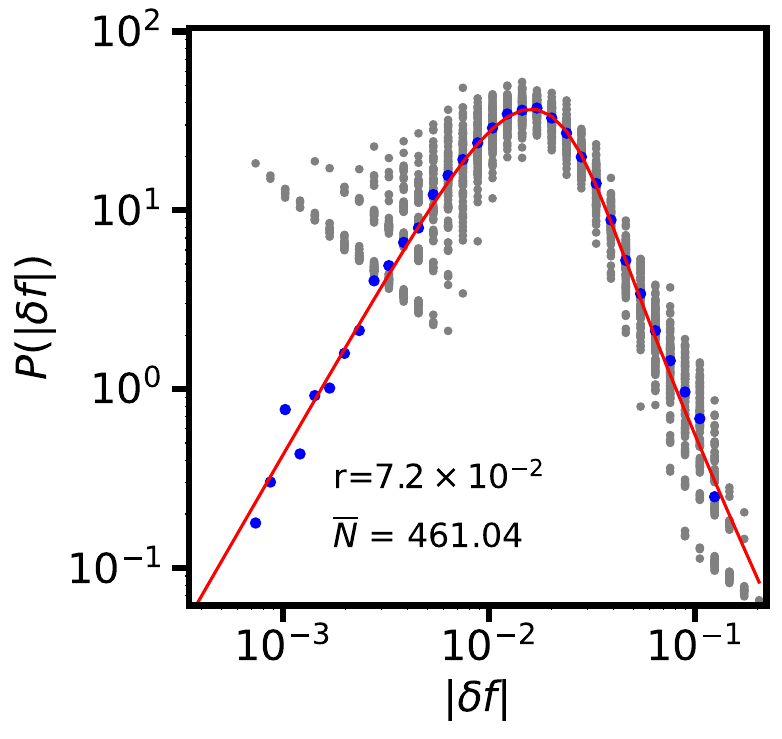} 
\includegraphics[width=0.33\textwidth]{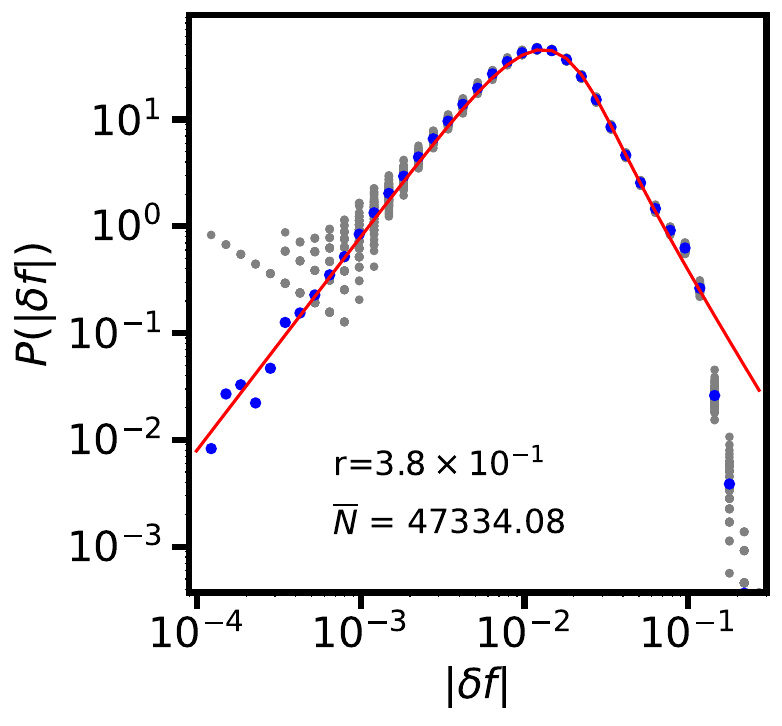} \\
\includegraphics[width=0.33\textwidth]{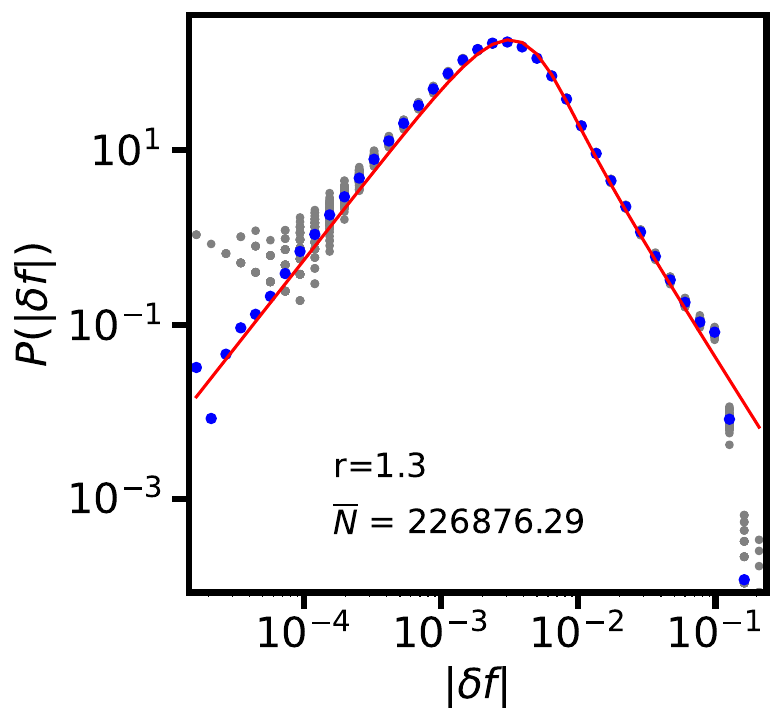} 
\includegraphics[width=0.33\textwidth]{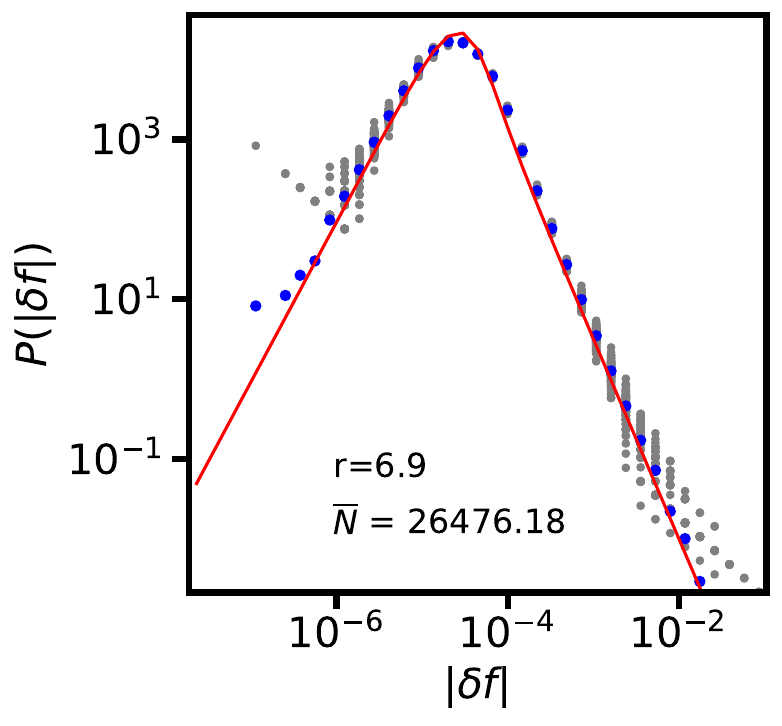} 
\includegraphics[width=0.33\textwidth]{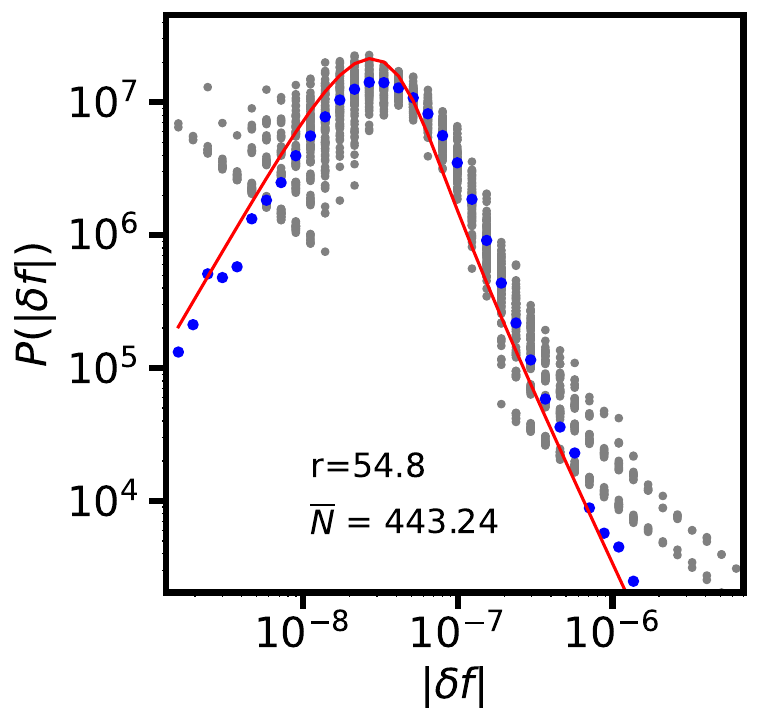} 
	\caption{Numerically evaluated distribution of force fluctuation $\mathcal{P}(\delta f)$ in different shells of increasing radius (clockwise from top left, $r=3.1\times 10^{-2}$, $7.2\times10^{-2}$, $0.38$, $1.3$, $6.9$ and 54.8 in units of $r_c$) for a Plummer model with $N=10^6$ and $10^2$ realizations (gray dots). The blue dots mark the PDF averaged over said ensemble while the red curve is the corresponding the Holtsmark distribution (\ref{holtsmark}) computed for the local value of the density $\rho(r)$ and average number of particles in the shell $\overline{N}$.}
	\label{force_fluctuation_histo}
\end{figure*}
Consistently with previous work, we accommodate the definition (\ref{holtsmark}) to the $\gamma-$models (\ref{dehnen}) and Plummer model (\ref{plummer}) using the local values of $\rho$ (e.g. see \citealt{1986SvAL...12..237P}) and use the typical local Maxwellian approximation\footnote{In general, the model appears to be more sensible to the specific form of $\mathcal{F}(\delta f)$ than that of $\Xi(v)$, see e.g. \cite{2018ApJ...867..163P,2020A&A...640A..79P} for the case where $m_t\approx 2m$.} for the velocity distribution of the model in the Chandrasekhar formula so that
\begin{equation}\label{fracvel}
\Xi(v)=(4\pi)^{-1}\left[{\rm Erf}\left(\frac{v}{\sqrt{2}\sigma}\right)-\sqrt{\frac{2}{\pi}}\frac{v}{\sigma}\exp\left(-\frac{v^2}{2\sigma^2}\right)\right].
\end{equation}
For the $\gamma-$models the radial velocity dispersion profile is given in integral form as
\begin{equation}
\sigma^2(r)=GMr^\gamma(r+r_c)^{4-\gamma}\int_r^\infty\frac{r^{\prime 1-2\gamma}{\rm d}r^\prime}{(r^\prime+r_c)^{7-2\gamma}}.
\end{equation}
In the two cases of interest explored here, $\gamma=0$ and 1 we have
\begin{equation}
\sigma^2(r)=\frac{GM}{30}\frac{r_c+6r}{(r_c+r)^2}
\end{equation}
and (see also \citealt{1990ApJ...356..359H})
\begin{align}
\sigma^2(r)=\frac{GM}{12r_c}\bigg[\frac{12r(r+r_c)^3}{r_c^4}\ln\bigg(\frac{r+r_c}{r}\bigg)+\nonumber\\
-\frac{r}{r+r_c}\bigg(25+\frac{52r}{r_c}+\frac{42r^2}{r_c^2}+\frac{12r^3}{r_c^3}\bigg)\bigg]
\end{align}
respectively, while for the Plummer sphere 
\begin{equation}
\sigma^2(r)=\frac{GM}{6\sqrt{r^2+r_c^2}}.
\end{equation}
The Coulomb logarithm appearing in Eq. (\ref{simpleDF}) is evaluated accounting for the exact integration over the impact parameter $b$ in hyperbolic orbits (see e.g. \citealt{binney}) as
\begin{equation}
\log\bigg[1+\bigg(\frac{b_{\rm max}}{b_{\rm min}}\bigg)^2\bigg]=\log\bigg\{1+\bigg[\frac{r_* \sigma^2}{G(m_t+m)}\bigg]^2\bigg\}\approx \log\Lambda.
\end{equation}
The choice of the minimum impact parameter $b_{\rm min}$ is made as usual to account for the angular momentum conservation in a $\varphi=\pi/2$ deflection at relative velocity of the order the local velocity dispersion $\sigma$, while $b_{\rm max}$ is set to $r_*$, a typical scale length of the order of the average inter-particle distance $\langle r\rangle=(8\pi mr_c^3/3M)^{1/3}\propto r_c/N^{1/3}$. We stress the fact that there is still not much of a consensus around the ''correct" choice of $b_{\rm max}$ (see \citealt{2018MNRAS.481.2041H}), with some authors assuming the scale length of the host $r_c$ while others (e.g. \citealt{1980PhR....63....1K,1983Ap&SS..97..435K}) preferring the local average interparticle distance $\overline{r}=(m/\rho)^{1/3}$ (see the discussion in \citealt{2020IAUS..351..532V}). Another choice could be the distance at which the correlations in the fluctuations of stellar density, velocity and gravitational potential induced by $m_t$ vanish, in a similar fashion to the Debye-Hu\"ckel theory in neutral plasmas (see \citealt{1968SvA....11..873M}). Ad hoc direct $N-$body simulations by \cite{2017PhRvE..96c2102M}, where $b_{\rm min}$ can be artificially fixed by the softening parameter $\epsilon$, seemed to indicate that $b_{\rm max}\approx r_c/3$. Here we note that, due to their rather low best resolution (i.e. $N\approx 1.7\times 10^4$), setting the maximum impact parameter to either $r_c/3$ or $\langle r\rangle$ inside the logarithm has little effect on the magnitude of the DF coefficient for resolutions below roughly $\sim 5\times10^{6}$.\\
\indent The stochastic equation of motion (\ref{langeq}) is solved by computing the dynamical friction term at all times as function of the local density and velocity distribution of the background model, and sampling the modulus of the force fluctuations from Eq. (\ref{holtsmark}) with a standard Monte Carlo. The direction of the fluctuations is assumed to be isotropic (i.e. $\delta\mathbf{f}$ is given a random direction by sampling a uniform distribution of angles). We integrated Eq. (\ref{langeq}) with the generalized leapfrog scheme implemented in \cite{2020A&A...640A..79P,2020IAUS..351...93D,2025arXiv250322479S} (for additional details and the original implementation see also \citealt{Mannella_2004,doi:10.1137/050646032}). The benefits of applying this integration technique are essentially the fact that it converges to the standard second-order Verlet scheme for vanishing noise and friction, and its rather fast convergence to asymptotic distributions for the case of simple Brownian motion. Other higher order and time-adaptive schemes, such as for example Runge-Kutta or Bulrisch-Stoer (see \citealt{doi:10.1137/050646032}), require longer computational times as dictated by smaller adaptive $\Delta t$ to obtain a comparable accuracy. It must be pointed out that the method introduced by \cite{Mannella_2004} in our implementation explicitly assumes a delta-correlated noise (i.e. subsequent encounters are independent), however, strictly speaking, it must be noted that the space-integrated force fluctuations are at zero-sum altogether, since $\delta f$ is related to density
fluctuations $\delta\rho$ through the usual Poisson equation. At constant total mass all fluctuations must integrate-out to zero. This implies that force fluctuations can not be completely uncorrelated, making the delta-correlated noise a rough approximation that could be, in principle, sharpened by introducing a correlation time $\tau_c$ (see e.g. \citealt{2025arXiv250602173T}). Other authors such as \cite{1999PhRvE..60.1567P,2003astro.ph.12434T}, used different forms of the correlation function of gravitational fluctuations, without finding significant qualitative differences with the delta-correlated noise for the case of strongly interacting systems. In some runs, where the $\delta\mathbf{f}$ term was artificially set to zero, we use instead a modified mid-point leapfrog, typically adopted for the integration of orbits with velocity dependent force terms (see for example \citealt{mik06}).
\subsection{Recovering the distribution of force fluctuations}
Using a locally tailored Holtsmark distribution in Eq. (\ref{langeq}) might seem a rather crude approximation. In order to check the reliability of this assumption, inspired by previous work by \cite{1973ApJ...179..885A,1974ApJ...188..469A} and \cite{1996A&A...305..999D,1996A&A...311..715D} we study the numerically evaluated distribution of the force (per unit mass) fluctuations for some discrete $N-$body realization of the models discussed in the previous Section.\\
\indent We define the force fluctuation at the position $\mathbf{r}_i$ of the $i$th particle in a discrete realization of the parent continuum system as
\begin{equation}
\delta\mathbf{f}(\mathbf{r}_i)=-\nabla\Phi(\mathbf{r}_i)+Gm\sum_{i\neq j =1}^N\frac{\mathbf{r}_i-\mathbf{r}_j}{||\mathbf{r}_i-\mathbf{r}_j||^3},
\end{equation}
where $\Phi$ is the smooth potential given by Eq. (\ref{dehnen}) or Eq. (\ref{plummer}), with normalization such that $M=Nm$. We then partition the model in $N_s$ logarithmically spaced radial shells and evaluate for each of them the probability density function $\mathcal{P}$ of the modulus of the force fluctuations $\delta f=||\delta\mathbf{f}||$ by particle counts as well as its average value $\langle\delta f\rangle$ and variance $\overline{\delta f}=\langle\delta f^2\rangle-\langle\delta f\rangle^2$.
 \begin{figure}
	\centering
	\includegraphics[width=\columnwidth]{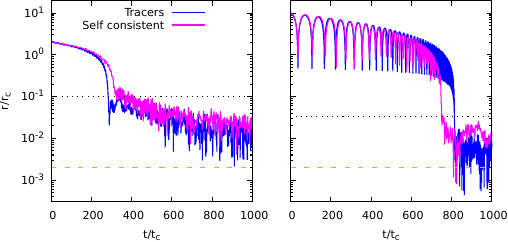}
	\caption{Radial decay of a massive tracer $m_t=10^{-3}M$ orbiting in a $\gamma=0$ (left) and a $\gamma=1$ (right) model in a self consistent $N-$body run (blue curves) and a simplified simulation where the same $N=10^6$ field particles move independently in the static smooth potential (magenta curves). The horizontal orange dashed line marks the value of the softening parameter $\epsilon=2\times 10^{-3}r_c$, while the black dotted line indicates the radius $r_{m_t}$ at which $M(r)=m_t$, equal to $\approx 0.1$ for the simulations with $\gamma=0$ and $\approx 0.033$ for those with $\gamma=1$.}
	\label{fig_test}
\end{figure}
\begin{figure}
	\centering
	\includegraphics[width=\columnwidth]{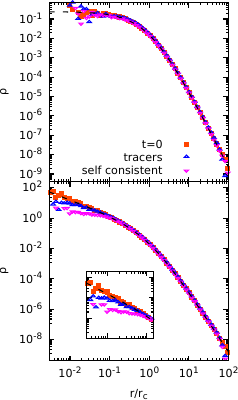}
	\caption{Initial density profile corresponding to a $\gamma=0$ (top) and a  $\gamma=1$ (bottom) model (filled orange squares) and corresponding density profiles at $t=10^3t_c$ in the self consistent and tracers in smooth potential simulations (blue upward triangles and magenta downward triangles, respectively). The black dashed lines mark $\rho(r)$ as given in Eq. (\ref{dehnen}). The points correspond to the same simulations of Fig. \ref{fig_test}.}
	\label{fig_rho}
\end{figure}
\begin{figure}
	\centering
	\includegraphics[width=0.9\columnwidth]{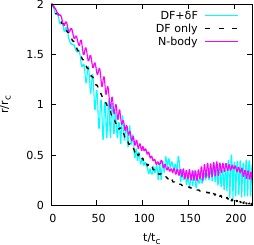}
	\caption{Decay of the galactocentric distance in a Plummer system for a particle with $m_t=2\times10^{-3}M$ in a $N-$body simulation with $N=10^6$ (magenta curve), and two semi analytical integrations in the parent smooth potential with dynamical friction and fluctuations (solid cyan line) and with dynamical friction only (black dashed line).}
	\label{fig_langevin}
\end{figure}
\begin{figure}
	\centering
	\includegraphics[width=0.9\columnwidth]{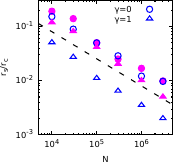}
	\caption{Core-stalling radius as function of $N$ for a particle of mass $m_t=10^{-3}M$ in a cored model ($\gamma=0$, circles) and a mildly cuspy model ($\gamma=1$, triangles). Filled magenta symbols refer to the full self consistent $N-$body simulations while empty blue ones to the simulations with non-interacting tracers. To guide the eye, the dashed line marks the $N^{-1/2}$ trend.}
	\label{fig_scaling}
\end{figure}
 In Fig.~\ref{plumer_average_variance} we show $\langle \delta f\rangle$ (left) and $\overline{\delta f}$ (right) for a Plummer model with $N=10^6$ for $N_s=20$ radial shells as function of their radius $r$. It appears that both average and the variance of the force fluctuations grow for decreasing $r$, being almost constant inside the core (i.e., below $r_c$). The associated histograms of $\mathcal{P}(\delta f)$ are shown in Fig.~\ref{force_fluctuation_histo} for 100 independent realizations (gray dots) and their average (blue dots), together with the Holtsmark distribution (red lines) computed evaluating numerically the integral in Eq. (\ref{holtsmark}) with the choice of $\alpha$ where $m=M/N$ and $\rho=\rho(r)$ for the specific model at hand, in this case given by Eq. (\ref{plummer}). Remarkably, the empirical force distribution\footnote{Notably, \cite{1962SvA.....5..809A} investigated the distribution of force fluctuations on a test particle surrounded by a finite number $\overline{N}$ of equal field particles finding rather good agreement with the Holtsmark distribution.} appears to agree rather well with the Holtsmark distribution, in particular for the shells enclosing larger fractions of $N$ for which the different realizations have a less pronounced scatter -the average number of field particles $\bar{N}$ in each shell is indicated at the bottom of the panels-, so that we can safely assume that the stochastic single particle simulations can be reasonably compared to honest direct $N-$body simulations, at least for what concerns the form of the fluctuation distribution. 
\section{Simulations and results}
\begin{figure*}
	\centering
	\includegraphics[width=0.95\textwidth]{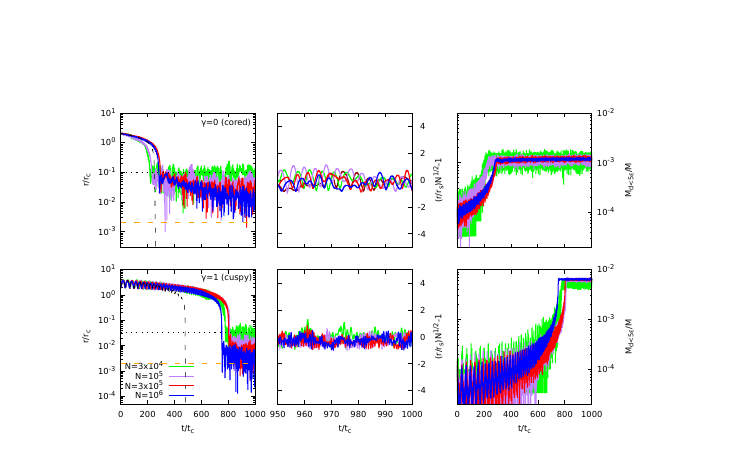}
	\caption{Decay of the galactocentric distance of a tracer particle of mass $m_t=10^{-3}M$ in models with $\gamma=0$ (top left panel) and 1 (bottom left) and the data collapse after a $N^{1/2}$ scaling (mid panels) to the time-averaged scaling radius $r_s$, and the evolution of the corresponding mass enclosed with in a sphere of radius $5\epsilon$ centered around $m_t$. The horizontal dotted and dashed lines have the same meaning as in Fig. \ref{fig_test}.}
	\label{fig_scaling_r}
\end{figure*}
\begin{figure}
	\centering
	\includegraphics[width=0.9\columnwidth]{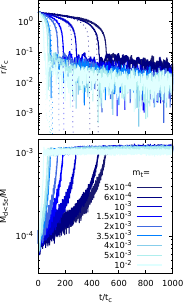}
	\caption{Decay of the galactocentric distance in a $\gamma=0$ cored system for test particle with masses in the range $2\times 10^{-4}\leq m_t/M \leq 10^{-2}$ and $N=10^6$ (top panel) indicated by increasingly lighter shades of blue, the dashed lines mark the radial decay predicted by the Chandrasekhar formula. Evolution of the mass enclosed with in a sphere of radius $5\epsilon$ centered around $m_t$ (bottom panel).}
	\label{fig_mass_ratio}
\end{figure}
\begin{figure}
	\centering
	\includegraphics[width=0.9\columnwidth]{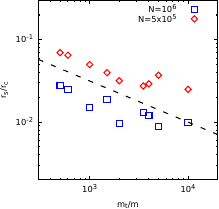}
	\caption{Stalling radius in a $\gamma=0$ model as a function of the mass ratio $m_t/m$ for the two system resolutions $N=5\times 10^5$ (red diamonds) and $N=10^6$ (blue squares). The dashed line marks the $(m_t/m)^{-1/2}$ trend.}
	\label{scalingmn}
\end{figure}
\begin{figure}
	\centering
	\includegraphics[width=0.9\columnwidth]{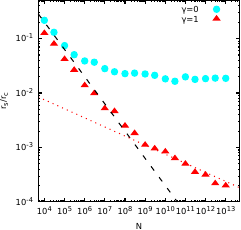}
	\caption{Core-stalling radius as function of $N$ for a particle of mass $m_t=10^{-3}M$ in a cored model ($\gamma=0$, cyan circles) and a mildly cuspy model ($\gamma=1$, red triangles) as extrapolated from Langevin simulations. As in Fig. \ref{fig_scaling} the dashed lines mark the $N^{1/2}$ best fit law in the range $10^4\leq N\leq 3\times 10^6$.}
	\label{fig_scalinglang}
\end{figure}
\subsection{Preliminary tests}
As a first set of numerical experiments we compared the orbital decay of the test mass $m_t$ in self consistent $N-$body simulations and non-interacting tracer systems, for different resolutions and initial orbital eccentricity. We find that, in both cases, independently on the specific model or details of the orbit the core-stalling can be always observed to some degree. For a given combination of initial orbital parameters and galaxy model the same mass $m_t$ systematically reaches a slightly smaller stalling radius $r_s$ in the simulation where it is slowed down by non-interacting tracers. In the simulations discussed hereafter, $r_s$ is estimated by time-averaging the radial position of $m_t$ over the last $10t_c$ of the simulation. Additionally we have also evaluated the PDF of the radial position over the same time span and extracted its mean and standard deviation, finding rather good agreement with the time average. Overall, the orbital decay is qualitatively comparable across the two types of runs, first affecting only $r_a$. As an example in Fig. \ref{fig_test} we show two test orbits with initial eccentricities of $e=0.001$ and 0.3, for a particle of mass $m_t=10^{-3}M$ propagated in a $\gamma=0$ and $\gamma=1$ models (left and right panel respectively). Both orbits plotted here stall at a $r_s$ larger than the softening length $\epsilon=2\times 10^{-3}r_c$ used in the simulation, here indicated by the horizontal dashed line. Typically, the orbit begins to stall (i.e. the decrease in pericentre becomes faster than that of the apocentre) once the enclosed mass fraction $M(r)$ becomes of the same order as $m_t$, as indicated by the black dotted lines. The critical radius $r_{m_t}$ where this happens encloses a systematically smaller number of particles for decreasing $N$. Therefore for the low resolution cases it may contain only order 10 particles. Further comparison between the two set of numerical experiments is discussed in Appendix \ref{ejconfronto}.\\
\indent As a means to explore the feedback of $m_t$ on the parent system, in Fig. \ref{fig_rho} we plot the radial density profile $\rho(r)$ at $t=10^3 t_c$ (triangles) against the initial one (squares), corresponding to the simulations shown in Fig. \ref{fig_test}. For the case of the cored system ($\gamma=0$, top panel), $\rho$ shows little to no evolution in both self consistent (blue upward triangles) and non-interacting tracer (magenta downward triangles) runs, small departure from the analytical profile around $r\approx 10^{-2}r_c$ can be rather unambiguously interpreted as resolution effects as they decrease for higher and increase for lower resolutions (here $N=10^6$). Test particles $m_t$ sinking in cuspy models appear to affect on a larger degree the density profile of the host system once reaching their inner regions, an effect that is related to the so-called BH scouring in elliptical galaxies hosting single or binary SMBHs (e.g. see \citealt{2021MNRAS.502.4794N} and references therein). In practice, strong scatters between stars on radial orbits and the massive BH deplete the stellar density profile thus reducing the slope of a possibly initially steeper stellar cusp. In the simulations performed here (see bottom panel of Fig. \ref{fig_rho}, and the related inset) we observe a substantial depletion of the central cusp of the $\gamma=1$ model in the simulation with non-interacting tracers, and an almost complete suppression of the cusp for the self-consistent case, where the background system becomes effectively ``cored'' as in the $N-$body simulations reported by \cite{2010ApJ...725.1707G}.\\
\indent Prompted by the good agreement between the theoretical Holtsmark distribution for an infinite system with the force fluctuation distribution in spherical shells on a finite model (cfr. Fig. \ref{force_fluctuation_histo}), it is natural to compare the orbital decay of a test mass in a $N-$body simulation with the trajectory of the same mass in the corresponding system modeled by Eq. (\ref{langeq}). We performed several runs for different values of $m_t/M$ and choices of the density profile $\rho$ finding that, in general, the stochastic equation method acceptably recovers both the time scale over which the mean galactocentric distance of given test mass halves and, in the case of a cored model its stalling radius $r_s$. As an example, Figure \ref{fig_langevin} shows the evolution of $r/r_c$ for a test particle $m_t=3\times 10^{-3}M$ initially placed on a nearly circular orbit at $r=2r_c$ in a Plummer model with $N=10^6$, propagated in a $N-$body simulation (magenta curve) and in a stochastic simulation (cyan curve). For comparison, we also added the orbital decay evaluated adding the Chandrasekhar DF (\ref{simpleDF}) to the smooth potential without fluctuations (black dashed line). In agreement with the literature discussed above, we observe that $m_t$ stalls at around $r_s\approx0.25 r_c$ with the $N-$body simulation always remaining in a sub-Chandrasekhar regime. Notably, at variance with the DF only integration, where the particle falls to the centre of the Plummer system remaining on approximately circular orbits of decaying radius, in both the $N-$body and the stochastic simulations some finite change in eccentricity (more marked for the orbit obtained integrating Eq. \ref{langeq}) can be appreciated. 
\subsection{$N-$body simulations}
We investigated the resolution-induced partial suppression of DF by evolving the same initial condition for the test particle $m_t$ in different realizations of $\gamma$ models with increasing number of particles $N$ and fixed total mass $M$.  In the interval $10^4\leq N\leq 3\times  10^6$, for each explored value of $m_t$ in the range between $2\times 10^{-4}$ and $10^{-2}$ in units of $M$, the stalling radius $r_s$ systematically settles at lower values for increasing $N$. We observe that, at fixed $N$ and $m_t$ and initial orbital parameters $e$ and $a$, the stalling radius is always lower in models with a central cusp, in both self-consistent and simplified $N-$body simulations with tracers.\\
\indent In Fig. \ref{fig_scaling} we show $r_s/r_c$ as a function of $N$ for cored and cuspy $\gamma-$models for $m_t=10^{-3}M$ initially placed on nearly circular orbits with semi-major axis $a\approx 2r_c$. We find that the stalling radius decreases with $N$ with a rather robust $1/\sqrt{N}$ trend, as indicated by the dashed line. The square root tendency is indicative of a discreteness effect-related mechanism behind the core-stalling (e.g. see \citealt{2014PhRvE..90f2910G,2019MNRAS.484.1456E} and references therein). This becomes even more evident if the evolution of the galactocentric distance of the different realizations is rescaled by the estimated value of $r_s$ times $N^{1/2}$. In Figure \ref{fig_scaling_r} we show the decay of $r_s$ up to $t=10^3t_c$ in a $\gamma=0$ (top) and $\gamma=1$ (bottom) model for the case of $m_t=10^{-3}M$ in a system of non-interacting tracers and the associated rescaled curves for the last $50t_c$ (left panels). A good data collapse is obtained, as also the amplitude of the fluctuations of $r_s$ fall on the same scale with a $N^{-1/2}$ trend (middle panels in Fig. \ref{fig_scaling_r}). A similar picture was also recovered for the self-consistent simulations (not shown here).\\
\indent Aiming at comparing with the idealized Chandrasekhar DF, we have also integrated Eq. (\ref{langeq}) for $m_t$ in the parent smooth models (i.e. setting to 0 the fluctuating term $\delta\mathbf{f}$). The black dashed curves in the left panels of Fig. \ref{fig_scaling_r} mark the corresponding predicted orbital decay. Notably, for the $\gamma=0$ cored case, the test particle might or might not experience the phase of super-Chandrasekhar DF reported by \cite{2006MNRAS.373.1451R} and \cite{2016MNRAS.455.3597Z}, depending on the specific value of $N$ in the simulation at hand. We observe that the onset of the super-Chandrasekhar regime happens at $N\lesssim 3\times 10^5$, suggesting that it might be connected with low resolution effects artificially enhancing the collisionality. The same effect is also observable in self-consistent simulations using comparable values of $\epsilon$. Interestingly, for all values of $N$ explored here the test particle stalls at $r_{m_t}>r_s\gtrsim \epsilon$. We performed additional tests at fixed $N$ and $m_t$ tuning the softening length (and the associated optimal $\Delta t$) between $3\times 10^{-5}r_c$ and $5\times10^{-2}r_c$, finding that for increasing $\epsilon$, the core stalling sets-in at later times and larger radii. According to the common wisdom, large values of the softening length render an $N-$body simulation ``less collisional'' as it depletes the large $\delta f$ tail of the two body force distribution by smoothening the pair interactions. At the same time, a large $\epsilon$ also hinders the DF by reducing the effective Coulomb logarithm (\citealt{2017PhRvE..96c2102M,2023MNRAS.519.5543M}).\\
\indent Once $m_t$ has stalled around $r_s$, the fraction of the host system's mass that surrounds it (i.e. the mass fraction enclosed in a fixed sphere centered on the instantaneous position of $m_t$) remains almost constant, even in the simulations without self gravity among the field particles. In the right panels of Fig. \ref{fig_scaling_r} we show as a function of time the mass fraction $M_{d<5\epsilon}/M$ inside a bona fide radius $d=5\epsilon$ (roughly of the same order of $r_{\rm inf}$ for $m_t=10^{-3}M$). For the cored model (top right panel), modulo some resolution dependent fluctuation, the value of $M_{d<5\epsilon}$ is practically the same for all $N$, while in the case of a cuspy system (bottom right panel), $M_{d<5\epsilon}$ at core stalling has the same value for $N\gtrsim 10^5$. In other words, since at fixed $m_t/M$ one has in our units approximately the same value of $r_{\rm inf}$ (cfr. Eq. \ref{rinflu}), this means that the test particle accretes the same amount of mass\footnote{We verified by checking particles indexes that the particles that ``dress'' $m_t$ are the same ones over the core stalling phase of the dynamics.} within its influence radius regardless of the specific resolution given by $N$. In practice, for increasing $N$ systematically larger number of orbits is trapped around $m_t$ that essentially becomes a ``dressed particle'', possibly contributing to a more significant erosion of the central density of cuspy systems via stronger field particles ejection (see Fif. \ref{fig_rho}).\\
\indent As expected, at fixed $M$ and $N$, increasing values of $m_t$ (marked by increasingly lighter shades of blue) result in lower stalling radii as shown in the top panel of  Fig. \ref{fig_mass_ratio} for $\gamma=0$, $N=10^6$ and values of the test particle mass in the range $2\times 10^{-4}\leq m_t/M \leq 10^{-2}$. Consistently with the predictions of the Chandrasekhar DF formula (represented in the figure by the dashed lines), heavier particles also sink in a shorter time. The mass enclosed within $5\epsilon$ around their position of $m_t$ (bottom panel of Fig. \ref{fig_mass_ratio}) settles in general to slightly lower values for larger $m_t$ (i.e. of the order of a 8\% difference between the largest and lowest values of the test mass considered here). This might appear in contradiction with what shown in Figure \ref{fig_scaling_r} (right panels) where at fixed $m_t$, $M_{d<5\epsilon}$ is practically independent on $N$ (and thus on $m_t/m$). However, one must notice that larger values of $m_t$ at fixed $m$, associated to lower $r_s$ might scatter more efficiently the field masses $m$, thus depopulating the control sphere at $d=5\epsilon$.\\
\indent In Figure \ref{scalingmn} we show the relation between the stalling radius and the ratio $m_t/m$ for two resolution $N=5\times 10^5$ (diamonds) and $N=10^6$ (squares). For both sets of points the trend is rather well described by $r_s=Cr_c\sqrt{m/m_t}$, where $C$ is a dimensionless constant of order unity. The stalling radius appears to have the same dependence on $m/m_t$ as the wander radius\footnote{ Actually, $N-$body simulations by \cite{2007AJ....133..553M} evidenced that the wander radius of a BH embedded in a $\gamma-$model might depend on its mass $m_{BH}$ with a much shallower power-law than that of Eq. (\ref{rwan}), as the rms of its Brownian velocity scales as $\langle V^2\rangle\propto m_{BH}^{-1/(3-\gamma)}$, thus implying a larger radial excursion even for large BH mass values, with respect to the predictions.} $r_{\rm wan}$ defined in Eq. (\ref{rwan}). We interpret this as another indication of the origin of the core stalling as a balance between friction and noise in a central potential, as originally envisaged for the derivation of $r_{\rm wan}$ with a Fokker-Planck approach (see again \citealt{1976ApJ...209..214B}). We noted that other studies, such as for example \cite{2006MNRAS.373.1451R}or \cite{2012MNRAS.426..601C} report a general trend of $r_s\propto r_c/2$, even in direct runs with resolution as large as $N=4\times10^6$. We conjecture that this apparent discrepancy with our numerical results could be related to their rather large softening length, ranging from $\approx r_c/3$ to $3r_c$ in some runs. We performed additional test runs at fixed $N$, $m_t$ and $\gamma$ (not shown here) finding that the stalling radius $r_s$ is systematically larger for increasing $\epsilon$.
\begin{figure*}
	\centering
	\includegraphics[width=0.9\textwidth]{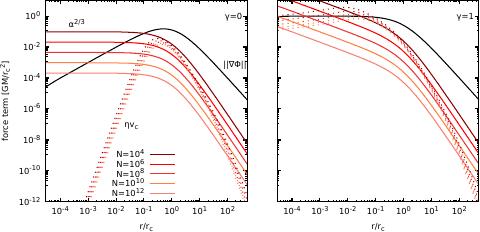}
	\caption{Radial dependence of the DF force acting on a particle of mass $m_t=10^{-3}M$ on a circular orbit at $v_c$ (dashed lines), of the typical force fluctuation $\overline{\delta f}\approx\alpha^{2/3}$ (solid lines, increasingly lighter shades of red correspond to larger values of $N$), and radial dependence of the strength of the gravitational field $||\nabla\Phi||$ (black heavy solid line). The left panel corresponds to the $\gamma=0$ case and the right one to the $\gamma=1$ case.}
	\label{fig_coeff}
\end{figure*}
\begin{figure*}
	\centering
	\includegraphics[width=0.9\textwidth]{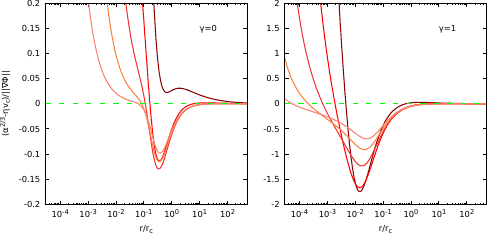}
	\caption{Radial dependence of the difference of between the DF term and typical force fluctuation, normalized by the local value of $||\nabla\Phi||$, for the same systems shown in Fig. \ref{fig_coeff}. The colour-coding of the curves is the same as in the previous figure.}
	\label{fig_coeffr}
\end{figure*}
\subsection{Stochastic simulations}
Aiming at exploring larger values of $N$ not accessible in direct $N-$body simulations we have carried out additional numerical experiments evolving the same initial orbits of Fig. \ref{fig_scaling} using Equation (\ref{langeq}) with resolutions up to $N=10^{13}$, roughly corresponding to the number of stars in a giant elliptical galaxy. Figure \ref{fig_scalinglang} presents $r_s/r_c$ as function of $N$ for the stochastic simulations with a $\gamma=0$ (circles) and $\gamma=1$ (triangles) parent galaxy model. Remarkably, in the interval of $N$ where we have also performed fully self-consistent $N-$body and simpler simulations with non-interacting field particles we recover the same $N^{-1/2}$ trend (dashed line) with similar values for $r_s/r_c$ to that reported in Fig. \ref{fig_scaling}. For the cuspy systems such trend extends up to $N\approx 10^9$, where for $m_t=10^{-3}M$ the amplitude of the fluctuations $\delta f$ stop balancing a significant fraction of the DF drag, then transitioning to a shallower $N^{-1/6}$, indicated by the brown dotted line. The origin of this slope change remains unclear. We speculate that it could be a numerical artifact related to the choice of $\Delta t$ in the solution of Eq. (\ref{langeq}). In the cored model the slope change happens at around $N\approx10^7$ leading to a seemingly $N-$invariant $r_s/r_c$ at fixed $m_t$ for values of $N$ larger than $10^9$. For larger values of $m_t$, $r_s$ systematically decreases for fixed values of $N$, while the slope changes move at lower $N$. It must be pointed out that, for values of the test mass larger than $\approx M/50$, for which in large $N$ models the effect of the force fluctuations would be essentially immaterial, the present linear approximation of the DF fails as $m_t$ would typically induce significant perturbations of both the local density and velocity distribution, well before sinking into the core. Unfortunately, checking via $N-$body the nature of the slope change between $N\approx 10^7$ and $10^9$ is made harder by the fact that the associated value of $r_s$ is around $\epsilon$ for $N\approx 10^8$, a radius that typically encloses for $\gamma=1$ a rather small number (order unity) of simulation particles, possibly enhancing the amplitude of the discreteness effects while being hardly coherent with the perturbed smooth potential approximation modelled by Eq. (\ref{langeq}).\\
\indent To illustrate the interplay between the contributions of the different force terms in Equation (\ref{langeq}) in Figure \ref{fig_coeff} we plot for different $N$ the radial dependence of the DF drag force acting on a mass $m_t=10^{-3}M$ on a circular orbit with velocity
\begin{equation}
v_{c}=\sqrt{\frac{GMr^{2-\gamma}}{(r+r_c)^{3-\gamma}}}
\end{equation}
and the typical amplitude of the force fluctuation $\overline{\delta f}\approx\alpha^{2/3}$, together with the radial gravitational field $||\nabla\Phi(r)||$. In the $\gamma=0$ cored model (left panel), the DF term drops significantly for $r\lesssim 0.1r_c$ as $v_c$ tends to 0 while $\sigma$ is nearly constant, resulting in a vanishing fractional velocity volume function (\ref{fracvel}). At the same time, for all values of $N$ the typical size of the force fluctuations becomes almost independent on $r$ as $\overline{\delta f}\propto\alpha^{2/3}$ and $\alpha\propto\rho$, with $\rho$ nearly constant below $r_c$ for a $\gamma=0$ model. The picture is qualitatively different in the $\gamma=1$ case (right panel), where the decrease of the DF force in the inner regions is way less pronounced and the amplitude of the fluctuating force increases as $1/r^{2/3}$. To further clarify how the force fluctuations compete with DF, in Fig. \ref{fig_coeffr} we plot for the same cases of Fig. \ref{fig_coeff} the difference of the amplitude of fluctuations and the DF force weighted over the local value of the gravitational field $||\nabla\Phi||$. For increasingly larger values of $N$ the ratio $(\alpha^{2/3}-\eta v_c)/||\nabla\Phi||$ becomes positive (i.e. force fluctuations mostly balance DF) at smaller radii in units of $r_c$ when $\gamma=1$ (right panel). In the $\gamma=0$ systems, for $N\gtrsim 10^9$, the radius at which the change of trend happens is almost $N-$independent (left panel), as evident from Fig. \ref{fig_scalinglang} where $r_s/r_c$ has little to no dependence on $N$ in that resolution range. We noted that such intersections between the curves in Fig. \ref{fig_coeffr} with the horizontal dashed line have the same trend with $N$ as $r_s/r_c$ in Fig. \ref{fig_scalinglang}. remarkably, for the $\gamma=1$ case no apparent slope change at around $N\approx 10^9$ was observed, further indicating a possible numerical origin for that feature. We recall that (see e.g. \citealt{1994hsmp.book.....G} Chap. 4), the characteristic time $\tau_{cr}$ in which a system described by Eq. (\ref{langeq}) relaxes is associated to the inverse of the friction coefficient $\eta$ (\citealt{1969ApJ...158L.139S}). The latter is essentially the first diffusion coefficient $D_f$ in the parent Fokker-Planck picture (see \citealt{1987degc.book.....S}). The other two coefficients $D_\perp$ and $D_\parallel$, associated to perpendicular and parallel heating (i.e. noisy fluctuations of the force) along the trajectory of $m_t$, respectively, are directly proportional to $m$ instead of to $m_t$ as for $D_f$. This implies that the times scale on which they effectively operate is typically larger for the cases with smaller values of $m_t/m$, as the dependence on $\rho$ and $\log\Lambda$ is the same as in $D_f$ (see the integral expression in \cite{2024MNRAS.534..957G}). We stress the fact that in the simulation presented here ($N-$body and stochastic) the integration time is systematically larger or comparable to the inverse of the noise coefficients below $r_c$.
\section{Outlook and conclusions}
Using $N-$body simulations and stochastic ODE methods we have investigated the dynamical friction acting on a heavy test mass in spherical systems characterized by a cored or cuspy density profile. Previous numerical studies of other authors have often reported a limited or in some cases substantial reduction of the gravitational drag in cored models, with respect to the estimate evaluated by the Chandrasekhar formula (\ref{simpleDF}). Typically, such discrepancy is interpreted as a consequence of the strong assumptions in the original formulation of dynamical friction in infinite and homogeneous systems. More recent semi-analytical work (\citealt{2021ApJ...912...43B,2022ApJ...926..215B}) on the other hand, indicates that dynamical friction in cores is naturally suppressed as its coefficient, dominated by the contribution of orbits in resonance with that of the test particle, vanishes towards the center with the vanishing fraction of nearly resonant particles, in a resolution-independent fashion.\\
\indent The main results of our work can be summarized as follows. The reduction of the dynamical friction force observed in $N-$body experiments appears to be resolution dependent (i.e. related to the number of simulation particles when fixing the normalization $G=M=r_c=1$) in the explored range $10^4\leq N\leq 3\times 10^{6}$ in both cored and cuspy models. The radius at which the test mass $m_t$ stalls without falling any further to the center has a marked $r_cN^{-1/2}$ dependence indicative of a underlying Poissonian fluctuation-driven process. Single particle stochastic integrations with analogous parameter reproduce the same trend, well beyond the $N$ values accessible by the direct simulations (Fig. \ref{fig_scalinglang}). We observe that, approximating the distribution of force fluctuations in said experiments with a Holtsmark function, accommodated to the local values of density and mean particle mass nicely recovers the the behaviour of orbits obtained by $N-$body simulations. Moreover, the histogram of such force fluctuations as obtained from $N-$particle realizations of a given system fits remarkably well the Holtsmark distribution over a large interval of radii. Comparing fully self consistent runs with others where the field particles do not interact with one another but propagate as independent tracers in the smooth potential of the model, revealed that in the range of $3\times 10^{-4}\lesssim m_t/M\lesssim 10^{-2}$ the contribution of a self gravitating wake does not enhance significantly the dynamical friction. The stalling radius determined from $N-$body simulations has a trend that is reasonably compatible $r_s\propto r_c/\sqrt{m_t/m}$.  This is reminiscent of what was obtained for the wander radius of black holes in dense stellar systems. It must be pointed out that \cite{2007AJ....133..553M}, in $N-$body experiments of BHs wandering in $\gamma$ models, using a simulation set-up essentially analogous to ours, determined that said $r_{\rm wan}$ in cusps broken below the influence radius $r_{\rm inf}$ could scale with $m_t$ with a much shallower power-law. This arises as the BHs' typical velocity at equilibrium was found to be $\langle v^2\rangle\propto m_t^{-1/(\gamma-1)}$ while the rms of its radial position is given by $\langle r^2\rangle\propto\sqrt{\langle v^2\rangle}$ when a linear restoring force is present (i.e. inside a nearly flatted out cusp the potential could be reasonably taken as harmonic). Nevertheless, we interpret the similarity of the trend in $r_s$ and (the expected one) in $r_{\rm wan}$ as an additional confirmation of the fact that core-stalling induced by a balance between friction and force fluctuations is the same phenomenon at the origin of a mass-dependent wander radius for a heavy particle starting at rest from the center of the potential well of a star cluster or galactic nucleus.\\
\indent From the stochastic simulations we extrapolate that in cuspy models the interplay between friction and fluctuation-related noise balances at smaller radii for increasing $N$ (up to a regime where $\delta f$ become virtually inexistent), while for cored models there is a critical size ($N_*\approx 3\times 10^7$ for $m_t\approx 10^{-3}M$) above which friction and noise counter each other seemingly independently on the resolution. We must point out that in the continuum limit (i.e. $N\rightarrow\infty$; $m\rightarrow 0$), in principle, both friction and noise should vanish leaving $m_t$ under the effect of $\Phi$ only. At variance with the analogous problems in plasma physics, where typically one deals with $N$ of the order of the Avogadro constant, a stellar system with $N\approx 10^{12}$ is still far from being considered continuum, and is still subjected to a certain degree to ``discreteness effect''.\\ 
\indent In conclusion, the results of the present work are not in contrast with the resonance-based and $N-$independent argument of \cite{2022ApJ...926..215B} that ascribes the core stalling to the vanishing friction coefficient in cores. Indeed, our findings reconcile the predicted $N-$independent cores stalling with the observed resolution dependent stalling radius $r_s$ reported by other $N-$body studies. In other words, for system sizes of the order of $N\sim10^3$ up to $10^6$ the radial interval over which both the friction and noise terms in Eq. (\ref{langeq}) are non-negligible with respect to the local smooth force field is significantly larger, regardless of the fact that the model is cuspy or cored. In cored models, for sufficiently small test mass $m_t$, the further outwards extent of this interval is larger than the critical radius at which the dynamical friction force on a nearly circular orbit begins to drop significantly. For increasing resolution (i.e. larger values of $N$ at fixed normalization and $m_t/M$), in cored models the $r-$independent fluctuations are smaller and counter dynamical friction nearly the same radius below which the friction coefficient has dropped of about two decades over less than a decade in radii (cfr. Figs. \ref{fig_coeff}\ref{fig_coeffr}). In the end, observing the onset of a resonance-driven core stalling in $N-$body experiments remains a challenging task, as serial direct numerical simulations are strongly limited in resolution. However, considering a test mass moving in a host system where field particles do not feel their mutual self gravity but orbit in a time-independent field is a problem that can be easily partitioned over different cores or graphic units allowing for a $N$ up to $10^{10}$. In this set-up, choosing a distribution function entirely constituted by circular orbits (that would likely be unstable when accounting for the self consistent forces) could serve as a computational test to probe a possible resonant driven suppression of the dynamical friction. 
\begin{acknowledgements}
  We thank Uddipan Banik, Frank van den Bosch, and Carlo Nipoti for the important discussions at different stages of this project. The anonymous Referee is also thanked for his/her valuable comments that helped improving the presentation of our results. This research was supported in part by grant NSF PHY-2309135 to the Kavli Institute for Theoretical Physics (KITP). PFDC wishes to thank the hospitality of the ''Laboratoire J.-A. Dieudonn\'e" at the university of C\^ote d'Azur where part of this work was completed. BM acknowledges support by the grant Segal ANR-19-CE31-0017 of the French Agence Nationale de la Recherche.
\end{acknowledgements}
\bibliographystyle{aa}
\bibliography{ras}
\begin{appendix}
\section{Analytical dynamical friction coefficient for the Plummer sphere}\label{plummerfriction}
\begin{figure}
	\centering
	\includegraphics[width=0.9\columnwidth]{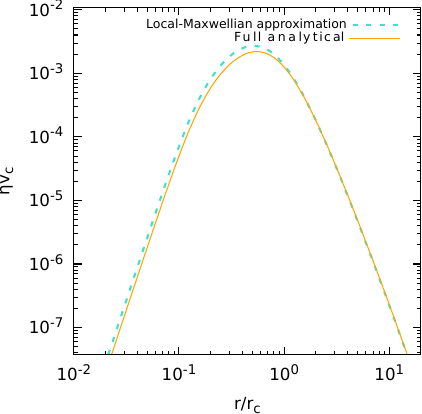}
	\caption{Radial profile of the dynamical friction force on a mass $m_t=10^{-3}M$ on a circular orbit in a Plummer model.}
	\label{locmax}
\end{figure}

The dynamical friction expressed in Eq. (\ref{simpleDF}) is often evaluated assuming that the phase space distribution of the underlying model can be factorized as $f(\mathbf{r},\mathbf{v})=\rho(\mathbf{r})f(\mathbf{v})$. In the commonly adopted local Maxwellian approximation, the position dependence of the velocity distribution $f(v)$ is restored by assuming a Maxwellian distribution with velocity dispersion proportional to $r$ via the Jeans equations for the associated density potential pair $(\rho,\Phi)$. In principle, in a spherically symmetric model with isotropic phase space distribution, the integral in Eq. (\ref{velfunct}) could be instead carried out over $f(\mathcal{E})=f(r,v)$. However, most often $f$ has a rather complicated form even for simple density profiles such as the $\gamma$-models. For the Plummer distribution, where $f(\mathcal{E})$ is given by Eq. (\ref{fplummer}) the integration becomes rather easy (e.g. see \citealt{2015ApJ...799...44M}) and one has
\begin{align}
\Xi^*(r,v)=4\pi 2^{3/2}k\Psi\int_0^{v(2\Psi)^{-1/2}}w^2(1-w^2)^{7/2}{\rm d}w=\nonumber\\
=4\pi 2^{3/2}k\Psi\int_u^\infty \frac{y^8}{(y^2+1)^6}{\rm d}w.
\end{align}
In the equation above $k=24\sqrt{2}r_c^2/7\pi^3G^5M^4$, $w=v/\sqrt{2\Psi}$ and $y=\sqrt{w^{-2}-1}$. Setting $u=\sqrt{2\Psi/v^2-1}$ one has
\begin{align}
\Xi^*(r,v)=\alpha\Psi^5\bigg[\frac{1}{2}\tan^{-1}(u)+\bigg(\frac{u^9}{2}-\frac{79}{21}u^7-\frac{64}{15}u^5-\frac{7}{3}u^3-\frac{u}{2}\bigg)\nonumber\\
\times(1+u^2)^{-5}-\frac{\pi}{4}\bigg],
\end{align}
where $\alpha=-7\pi\sqrt{2}k/16$.\\
\indent In Fig. \ref{locmax} we show the radial profile of the dynamical friction force acting on a test mass $m_t=10^{-3}M$ moving on a circular orbit at the local circular velocity $v_c=\sqrt{GMr(r^2+r_c^2)^{-3/2}}$, evaluated in the local Maxwellian approximation (dashed lines) and integrating over the distribution function (solid line), so that the term $\rho\Xi(v)$ in Equation (\ref{simpleDF}) is substituted by $\Xi^*(r,v)$ defined above. At $r\gg r_c$ the two expressions do not give appreciably different drag force, while below $r\approx5\times 10^{-1}r_c$, the local Maxwellian approximation overestimates the dynamical friction of a factor about 1.4, we conclude that using such approximate form for the friction force does not alter significantly its effect, in particular since the region where the error is greater corresponds to a radial range where its contribution itself becomes under-dominant. 
\section{Further comparison between self-consistent and tracer simulations}\label{ejconfronto}
\begin{figure}
	\centering
	\includegraphics[width=0.9\columnwidth]{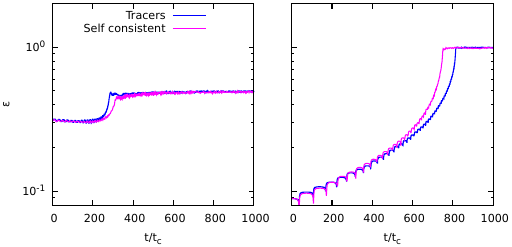}
	\caption{Evolution of the relative orbital energy per unit mass $\mathcal{E}$ for a test mass $m_t=10^{-3}$ in a $\gamma=0$ (left) and $\gamma=1$ (right) model. In both cases $N=10^6$ and the blue and magenta curves refer to the tracer in smooth potential and self-consistent runs, respectively.}
	\label{confe}
\end{figure}

\begin{figure}
	\centering
	\includegraphics[width=0.9\columnwidth]{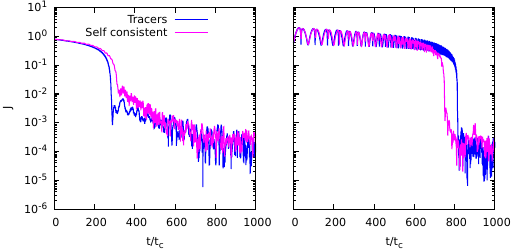}
	\caption{Evolution of the norm of the orbital angular momentum $J$ for the same systems of Fig. \ref{confe}.}
	\label{confj}
\end{figure}

 As the asymptotic stalling radius is roughly the same in self-consistent simulation and simplified smooth potential and tracer ones, we expect that the (relative) energy of $m_t$ relaxes to comparable values once the its radial coordinate has settled at $\approx r_s$. In Fig. \ref{confe} we show the evolution of $\mathcal{E}$ for the same runs of Fig. \ref{fig_test}. In this case, where the resolution is $N=10^6$ we observe a remarkably good agreement of the final values of the relative energy for both values of $\gamma$ with relatively small fluctuations. Smaller values of $N$ (not shown here) also yield rather similar values, though bearing larger fluctuations around the final value for $N\lesssim10^5$. By contrast, the norm of the angular momentum $J$ shown for the same simulations in Fig. \ref{confj} shows wilder fluctuations (even at $N$ as large as $3\times 10^6$), of about the 10\% of its asymptotic value. We interpret this as a result of the more chaotic behaviour of orbits in the inner region of both models that reflects in violent changes in orbital eccentricity and inclination while the energy remains nearly constant.

\end{appendix}
\end{document}